\documentclass[a4paper,12pt]{article}
\usepackage{verbatim} 
\usepackage{jheppub} 

\usepackage[T1]{fontenc} 
\usepackage[utf8]{inputenc}
\usepackage{graphicx}
\usepackage{tikz}
\usepackage{import}
\usepackage{accents}
\usepackage{mathrsfs,amsmath,amssymb,slashed}
\usepackage{multirow,multicol}

\newcommand{\dd}{{\mathrm{d}}}
\newcommand{\newg}{{\ell}}

\title{2D holography beyond the Jackiw--Teitelboim model}

\author[a]{Florian Ecker,}
\emailAdd{eckerflorian32@gmail.com}
\author[b]{Carlos Valc\'arcel}
\emailAdd{valcarcel.flores@gmail.com}
\author[c]{and Dmitri Vassilevich}
\emailAdd{dvassil@gmail.com}

\affiliation[a]{Institute for Theoretical Physics, TU Wien, Wiedner Hauptstr.~8-10/136, A-1040 Vienna, Austria}
\affiliation[b]{Instituto de F\'isica - Universidade Federal da Bahia, C\^ampus Universit\'ario de Ondina, 40210-340, Salvador, B.A. Brazil}
\affiliation[c]{CMCC-Universidade Federal do ABC, Santo Andr\'e, S.P. Brazil}
\affiliation[c]{Department of Physics, Tomsk State University, Tomsk, Russia}

\abstract{Having in mind extensions of 2D holography beyond the Jackiw--Teitelboim model we propose holographic counterterms and asymptotic conditions for a family of asymptotically AdS$_2$ dilaton gravity models leading to a consistent variational problem and a finite on-shell action. We show the presence of asymptotic Virasoro symmetries in all these models. The Schwarzian action generates (a part) of the equations of motion governing the asymptotic degrees of freedom. We also analyse the applicability of various entropy formulae. By a dilaton-dependent conformal transformation our results are extended to an even larger class of models having exotic asymptotic behavior. We also analyse asymptotic symmetries for some other classes of dilaton gravities without, however, constructing holographic counterterms.}

\begin{document}

\maketitle

\section{Introduction}\label{sec:Intro}
The interest in the holographic correspondence in two dimensions was boosted by the discovery \cite{Maldacena:2016hyu,Maldacena:2016upp} that Jackiw--Tetelboim (JT) gravity \cite{Jackiw:1984,Teitelboim:1983ux} is dual to the Sachdev--Ye--Kitaev (SYK) model \cite{Sachdev:1992fk,Kitaev}. This correspondence has  as an intermediate point the Schwarzian action which appears as a semiclassical limit of the SYK model and as an effective boundary theory in JT gravity. A natural question which arises is: How far can this or similar correspondence schemes be extended to generic 2D dilaton gravity models?

Despite their great diversity all dilaton gravities in 2D share some fundamental properties. They are all topological, do not have local degrees of freedom, and all their classical solutions may be written in a closed (and rather simple) way \cite{Grumiller:2002nm}. Moreover, even quantum dilaton gravities without matter are locally quantum trivial \cite{Kummer:1996hy}. The whole dynamics in these models is concentrated on the boundaries or asymptotic regions. Thus, general dilaton gravities seem to be natural candidates for testing the holographic principle. JT gravity is not special in this respect. 

Of course, the holographic correspondence in  dilaton gravity models other than JT received a lot of attention over the recent years. Even not taking into account the extensions by adding more fields (SUSY, higher spins, etc.) we have to mention the following interesting and important developments. First of all, we note that the argumentation by Jensen \cite{Jensen:2016pah} is of rather general nature and is not restricted to just JT.
The dualities to matrix models proposed \cite{Saad:2019lba} for the JT model were extended to more general dilaton gravities \cite{Witten:2020wvy} (see also \cite{Momeni:2020tyt,Johnson:2020lns,Turiaci:2020fjj} for related study). Rather general dilaton gravities were encountered in the context of $T\bar T$ deformations \cite{Gross:2019ach,Ishii:2019uwk,Grumiller:2020fbb} or if AdS$_2$ holography is approached from higher-dimensional theories \cite{Blake:2016jnn,Cvetic:2016eiv,Hong:2019tsx,Narayan:2020pyj}. Thermodynamical aspects of the holographic correspondence with deformed JT gravities were addressed in \cite{Cao:2021upq}. A flat space holographic correspondence was realized in \cite{Afshar:2019axx} for a conformally transformed Callan--Giddings--Harvey--Strominger (CGHS) model \cite{Callan:1992rs} (also called a flat space JT model in some literature). $R^2$ corrections to AdS$_2$ holography were considered in \cite{Aniceto:2020saj}. An overview and some further references may be found in \cite{Trunin:2020vwy}.

While speaking about the holographic correspondence we would like to stress one important aspect of this correspondence which was neglected in most of the papers cited above. We mean a set of asymptotic conditions together with a boundary action including all holographic counterterms which ensure the consistency of the variational problem and a finite total on-shell action. For pure dilaton gravities, such holographic counterterms are only known in the JT model with rather general asymptotic conditions \cite{Grumiller:2017qao} and conformally transformed CGHS asymptotically flat model \cite{Afshar:2019axx}.\footnote{For the sake of completeness, we would like to mention the boundary actions obtained for supersymmetric \cite{Cardenas:2018krd}, and higher spin \cite{Gonzalez:2018enk,Afshar:2020dth} extensions of 2D dilaton gravities and for specific limits \cite{Grumiller:2020elf,Gomis:2020wxp}  of the JT model.} For generic dilaton gravities, the full boundary action is known only in the case of a non-fluctuating dilaton \cite{Grumiller:2007ju} which does not lead to interesting asymptotic symmetries. 

In this work, we extend the list of bulk theory candidates for 2D holographic correspondence. We first take a one-parameter family of asymptotically AdS$_2$ dilaton gravities and find there a set of asymptotic conditions and a boundary action leading to a consistent variational problem. We find that the asymptotic symmetry algebra for these asymptotic conditions is the Virasoro one. By a dilaton-dependent conformal redefinition of the metric we extend these results to an even wider class of models which however have rather exotic asymptotic behavior ("asymptotically trumpet", for example). We show, that a Schwarzian action generates a part of the equations of motion for the boundary degrees of freedom. To gain some information on CFT duals, we check the Cardy-type entropy formulae. The entropy appears to be inconsistent with the standard Cardy formula, but is reproduced in a modified version. Lastly, we define asymptotic symmetry algebras for an even larger class of models and asymptotic conditions, however without analysing the consistency with the boundary action. Omitting this technically most cumbersome step makes these last result less reliable but allows to show the directions for future studies.

This paper is organized as follows. In the next section, we introduce generic dilaton gravities in first and second order formulations and define a particular family of asymptotic AdS$_2$ models which will be the primary object of our study. The asymptotic conditions are proposed in section \ref{sec:asc} and in section \ref{sec:asa} the Virasoro asymptotic symmetry algebra is found. In section \ref{sec:boa} we propose a boundary action and show its consistency with the asymptotic conditions. Various entropy formulae are analysed in section \ref{sec:entropy}. In section \ref{sec:Sch} the equation of motion for the boundary degrees of freedom is related to a Schwarzian action. In section \ref{sec:con} we further extend the class of models by means of a dilaton-dependent conformal transformation. The symmetries of a large family of asymptotic conditions are found in section \ref{sec:more}. Some discussion of the results is contained in section \ref{sec:con}. Exact classical solutions of all dilaton gravities in Euclidean signature are described in Appendix \ref{sec:ex}.

\section{A family of asymptotically $\mathrm{AdS}_2$ dilaton gravities}

\subsection{Second-order formulation}
Classical and quantum properties of 2D dilaton gravities have been reviewed in \cite{Grumiller:2002nm}. With sufficient degree of generality, dilaton gravities on a 2-dimensional Euclidean manifold $\mathcal{M}$ are described by the action
\begin{equation}\label{D01}
I_{\mathrm{2nd}} = -\frac{1}{2} \int_{\mathcal M}\mathrm{d}^{2}x\:\sqrt{g}\left[\Phi R-U\left(\Phi\right)\left(\partial\Phi\right)^{2}-2V\left(\Phi\right)\right]
\end{equation}
depending on the metric $g_{\mu\nu}$ and the dilaton field $\Phi$. The curvature scalar is denoted by $R$ and $U$, $V$ are two arbitrary potentials.\footnote{The majority of models which appeared so far in the context of 2D holographic correspondence had $U=0$.} Different choices of potentials describe different kinds of dilaton gravity models. 

There is a particularly important family \cite{Katanaev:1997ni} of these models parametrized by two real numbers $a$ and $b$ together with a scale parameter $B$:
\begin{equation}\label{ab01}
U\left(\Phi\right)=-\frac{{a}}{\Phi},\;\;\;\;\;\;\; V\left(\Phi\right)=-\frac{B}{2}\Phi^{{a}+{b}}.
\end{equation}
This family includes the Jackiw-Teitelboim model $(a=0,b=1)$ \cite{Teitelboim:1983ux,Jackiw:1984}, the CGHS model $(a=1,b=0)$ \cite{Witten:1991yr,Callan:1992rs} and the spherically reduced gravity models \cite{Thomi:1984na,Hajicek:1984mz}. The classical solutions for dilaton gravities in the $(a,b)$-family and their properties were thoroughly studied in \cite{Katanaev:1997ni,Liebl:1997ti}. 

The line
\begin{equation}
a+b=1 \label{abAdS}
\end{equation}
in the $(a,b)$ plane corresponds to the models with an $\mathrm{AdS}_2$ ground state. Global solutions in this subfamily were found by Lemos and S\'a \cite{Lemos:1994py}. As noticed in \cite{Liebl:1997ti}, static metrics with proper asymptotic behaviour can be constructed for $0\leq a <2$ only. The point $(a,b)=(1,0)$ actually corresponds to a vanishing scalar curvature and is very specific in many respects. Thus we shall restrict ourselves to the interval $a\in [0,1)$ which is connected to the JT model $a=0$. For technical reasons, we shall move the upper boundary of this interval to $a=2/3$, see a remark at the end of section \ref{sec:boa}. Some aspects of the holographic correspondence in this family of models were considered as early as in \cite{Cadoni:2001ew}. However, important ingredients such as a boundary action consistent with asymptotic conditions were missing in that work.

\subsection{First-order formulation}

Instead of working in the second-order formulation (\ref{D01}), we can use the Cartan variables (zweibein and spin-connection) along with two auxiliary scalar fields. The introduction of these new fields makes the field equations first order differential equations which simplifies the analysis. The first-order formulation which is described by the action
\begin{equation}\label{D02}
I_{\mathrm{1st}}=\int_{\mathcal M}\left[Y^{\alpha}\mathrm{d}e_{\alpha}+\Phi\mathrm{d}\omega+\epsilon^{\alpha\beta}Y_{\alpha}\omega\wedge e_{\beta}-\frac{1}{2}\epsilon^{\alpha\beta}\mathcal{V} e_{\alpha}\wedge e_{\beta}\right]
\end{equation}
with
\begin{equation}
\mathcal{V}=\tfrac 12 U(\Phi) Y^\gamma Y_\gamma + V(\Phi).\label{calV}
\end{equation}
Here, the Greek indices $\alpha,\beta,\gamma \in \{ 1,2\}$ are raised and lowered with the Euclidean flat metric $\delta_{\alpha\beta}=\mathrm{diag}(1,1)$, $e_\alpha$ is a zweibein one-form, $\omega$ is a connection one-form. The antisymmetric Levi--Civita symbol is denoted by $\epsilon^{\alpha\beta}$ and, by convention, $\epsilon^{12}=1$. $Y^\alpha$ is an auxiliary scalar field which generates the (generalized) torsion constraint.

To prove that (\ref{D02}) is equivalent to (\ref{D01}) one has to eliminate the fields $Y^\alpha$ by means of their algebraic equations of motion and re-express the action in terms of the metric instead of the zweibein, see \cite{Grumiller:2002nm} for details. In the course of these manipulations an integration by parts is needed, so that
\begin{equation}
I_{\mathrm{2nd}}=I_{\mathrm{1st}} - \int_{\partial\mathcal{M}} \Phi\, t ,\label{I21}
\end{equation}
where 
\begin{equation}
t=U(\Phi)\, Y^\alpha e_\alpha \label{deft}
\end{equation}
is a contorsion one-form. From \eqref{I21} we notice that the difference between the second and first-order actions is only a boundary term, this proves the equivalence at a classical level. Another interesting consequence is that, if the on-shell first-order action is zero, the contribution of its second-order equivalent is a pure boundary term. In fact, in Appendix \ref{sec:ex} we show that dilaton theories with linear potential $V(\Phi)$ and inverse linear $U(\Phi)$ fall into this category. This result will be useful later, when we compute the on-shell action. 

Without loss of generality the scale factor $B$ can be fixed to take any convenient value. We take $B=2-a$, as in \cite{Grumiller:2007ju}. Thus, the models considered in this work correspond to the potentials
\begin{equation}
U(\Phi)=-\frac a\Phi \,,\qquad V(\Phi)=-\frac{(2-a)\Phi}2\,.\label{UV}
\end{equation}
The corresponding equations of motion read
\begin{eqnarray}
\dd \Phi + \epsilon^{\alpha\beta}Y_\alpha e_\beta &=& 0,\label{eom1}\\
\dd Y^\alpha +\epsilon^{\alpha\beta} \left[ Y_\beta \Omega + \frac a{2\Phi} \left( Y^\gamma Y_\gamma e_\beta -2 Y_\beta Y^\gamma e_\gamma \right) + \frac{2-a}2 \Phi\, e_\beta \right] &=& 0, \label{eom2}\\
\dd\Omega +\frac 14 \left[ (1-2a)(2-a) + a \Phi^{-2}\, Y^\alpha Y_\alpha \right]\, \epsilon^{\beta\gamma} e_\beta \wedge e_\gamma&=& 0,\label{eom3}\\
\dd e^\alpha + \epsilon^{\alpha\beta}\Omega\wedge e_\beta &=& 0\label{eom4},
\end{eqnarray}
where $\Omega=\omega-t$ is the Levi-Civita connection.

As a consequence of these equations the quantity
\begin{equation}\label{Cafam}
\mathcal{C}=\frac 12 \left( -\Phi^{2-a} + Y^\alpha Y_\alpha\, \Phi^{-a} \right)
\end{equation}
is absolutely conserved, $\dd\mathcal{C}=0$ (cf. (\ref{S02}) and (\ref{eQw})). In the Poisson-sigma language, $\mathcal C$ is the Casimir function and it is related to the mass of the black hole solutions. 

Exact solutions for these models are presented in Appendix \ref{sec:ex}. The curvature scalar depends on the dilaton and reads 
\begin{equation}\label{ab03}
R=-2\mathcal{C}{a}\Phi^{-\left(2-{a}\right)}-2\left(1-{a}\right)^{2}.
\end{equation}
One can see, that independently of $\mathcal{C}$ the region $\Phi\to\infty$ is asymptotically (Euclidean) AdS$_2$. 

\section{Asymptotic conditions}\label{sec:asc}
The asymptotic conditions have to be consistent with the equations of motion. Thus,
to formulate asymptotic conditions we have to solve the equations of motion\footnote{Alternatively, one may try to apply suitable coordinate and gauge transformations to the exact solutions from Appendix \ref{sec:ex}. This however appears to be technically even more involved.} in the asymptotic region. It is convenient to change the variables as follows
\begin{equation}\label{ab07}
\Phi\equiv\tilde{\Phi}^{\frac{2}{2-{a}}},\;\;\; Y^\alpha =\tilde{\Phi}^{\frac{{a}}{2-{a}}}X^\alpha .
\end{equation}
Then, the conserved quantity (\ref{Cafam}) becomes
\begin{equation}\label{ab08}
\mathcal{C}=\frac{1}{2}\left( -\tilde{\Phi}^{2}+X^\alpha X_\alpha \right).
\end{equation}
The equations of motion (\ref{eom1}) - (\ref{eom4}) take the form
\begin{eqnarray}
\frac 2{2-a}\dd \tilde\Phi + \epsilon^{\alpha\beta}X_\alpha e_\beta &=& 0,\label{eom11}\\
\dd X^\alpha +\epsilon^{\alpha\beta} \left[ X_\beta \Omega - \frac a{2\tilde\Phi} X_\beta X^\gamma e_\gamma + \frac{2-a}2 \tilde\Phi\, e_\beta \right]&=& 0, \label{eom12}\\
\dd \Omega + \frac 12 \left[ (1-a)^2 +a \tilde\Phi^{-2}\mathcal{C}\right] \epsilon^{\beta\gamma}e_\beta\wedge e_\gamma &=& 0\,.\label{eom13}
\end{eqnarray}
Eq.\ (\ref{eom4}) remains unchanged.

We suppose that the asymptotic line element on our two-dimensional manifold $\mathcal M$ has the form
\begin{equation}\label{ab10}
\mathrm{d}s^{2}=\mathrm{d}\rho^{2}+h(\rho,\tau)^{2}\mathrm{d}\tau^{2},
\end{equation}
where $\rho\to\infty$ is the asymptotic region and $\tau$ plays the role of a coordinate along the asymptotic boundary, we also assume periodicity $\tau\rightarrow \tau + \beta_\tau$. The zweibein fields may be chosen as
\begin{equation}
e_{\tau 1}=h,\qquad e_{\rho 2}=1 \,.\label{ee}
\end{equation}
Eqs.\ (\ref{ab10}) and (\ref{ee}) are in effect gauge conditions which need to be imposed in the asymptotic region only.

Other choices of gauge will be considered in section \ref{sec:more}. For those choices, we shall be able to find the asymptotic symmetry algebra, though the full analysis including the boundary action will be too complicated.

Our next step is the construction of asymptotic solutions to the equations (\ref{eom11}) - (\ref{eom13}) containing the derivatives with respect to the radial coordinate $\rho$. The possibility to separate the $\rho$- and $\tau$- equations of motion is one of the advantages of the first-order formalism. Eq. (\ref{eom4}) yields
\begin{equation}
\Omega_{\rho}=0,\qquad \Omega_{\tau}=\partial_{\rho}h \label{SolOm}
\end{equation}
Eq. (\ref{eom11}) is equivalent to a pair of equations,
\begin{eqnarray}
X^{1} &=& -\frac{2}{2-{a}}\partial_{\rho}\tilde{\Phi}   \label{ab11a}\\
X^{2} &=& \frac{2}{2-{a}}h^{-1}\partial_{\tau}\tilde{\Phi}  . \label{ab11b}
\end{eqnarray}
The $\dd\rho$-component of Eq.\ (\ref{eom12}) reads
\begin{eqnarray}
\partial_{\rho}X^{1}+\frac{\left(2-{a}\right)}{2}\tilde{\Phi}-\frac{{a}}{2}\tilde{\Phi}^{-1}\left(X^{2}\right)^{2}	 &=&	0 \label{ab13a}\\
\partial_{\rho}X^{2}+\frac{{a}}{2}\tilde{\Phi}^{-1}X^{2}X^{1} &=& 0. \label{ab13b}
\end{eqnarray}
The second of these equations, Eq.\ (\ref{ab13b}), can be solved jointly with (\ref{ab11a}) yielding
\begin{equation}
X^2=\newg\left(\tau\right)\tilde{\Phi}^{\frac{{a}}{2-{a}}},\label{eom21}
\end{equation}
where $\newg$ is arbitrary function of $\tau$. 

By substituting \eqref{ab11a} in \eqref{ab13a} we obtain
\begin{equation}\label{ab14}
\partial_{\rho}^{2}\tilde{\Phi}-\frac{\left(2-{a}\right)^{2}}{4}\tilde{\Phi}
+\frac{\left(2-{a}\right){a}}{4}\, \tilde{\Phi}^{-\frac{2-3{a}}{2-{a}}}\newg^{2}=0.
\end{equation}
This equation has to be solved perturbatively with the third term playing the role of perturbation. The main part is solved by $\tilde\Phi_0=x_R(\tau)\exp(\tfrac 12 (2-a)\rho) +x_L(\tau)\exp(-\tfrac 12 (2-a)\rho)$ with two arbitrary functions $x_{R,L}(\tau)$. We substitute $\tilde\Phi_0$ back to (\ref{ab14}) to obtain
\begin{equation}
\tilde{\Phi}\left(\rho,\tau\right) \simeq x_{R}\left(\tau\right)e^{\frac{1}{2}\left(2-{a}\right)\rho}+x_{L}\left(\tau\right)e^{-\frac{1}{2}
\left(2-{a}\right)\rho}+\frac{\left(2-{a}\right)}{8\left(1-{a}\right)}
\newg^{2}x_{R}^{-\frac{2-3{a}}{2-{a}}}e^{-\frac{1}{2}\left(2-3{a}\right)\rho}.\label{bc01a}
\end{equation}
The next order correction to this equation is $\mathcal{O}(e^{-\frac 12(6-7a)\rho})$. This correction has an extra factor of $e^{- (2-3a)\rho}$ as compared to the smallest term on the right hand side of (\ref{bc01a}) (which is the second term). Thus, the correction is relatively small if $a<2/3$. Higher order corrections follow the same pattern and are suppressed for $a$ in the same range.

The equation for $h$ 
\begin{equation}\label{ab12}
\partial_{\rho}^{2}h-\left[\left(1-{a}\right)^{2}+{a}\tilde{\Phi}^{-2}\mathcal{C}\right]h=0
\end{equation}
follows from (\ref{eom13}). It is also solved perturbatively. We assume that the $\rho$-equations hold. This guarantees that $\partial_\rho \mathcal{C}=0$, but does not yet imply that $\mathcal{C}$ is a constant in $\tau$.
\begin{equation}\label{bc02}
h\simeq e^{\left(1-{a}\right)\rho}+\mathcal{L}e^{-\left(1-{a}\right)\rho}+
\frac{1}{2-{a}}x_{R}^{-2}\mathcal{C}e^{-\rho}.
\end{equation}
The corrections here are $\mathcal{O}(e^{-(3-a)\rho})$ and thus are suppressed in the asymptotics for any $a\in [0,1)$.

Asymptotically, the solutions of $\rho$-equations for $X^1$ and $X^2$ are
\begin{eqnarray}
X^{1}&\simeq&-x_{R}e^{\frac{1}{2}\left(2-{a}\right)\rho}+x_{L}e^{-\frac{1}{2}\left(2-{a}\right)\rho}
+\frac{\left(2-3{a}\right)}{8\left(1-{a}\right)}\newg^{2}x_{R}^{-\frac{2-3{a}}{2-{a}}}
e^{-\frac{1}{2}\left(2-3{a}\right)\rho},\label{bc01b}\\
X^{2} &\simeq& \newg x_{R}^{\frac{{a}}{2-{a}}}e^{\frac{{a}}{2}\rho}\left[1+\frac{{a}}{2-{a}}
x_{R}^{-1}x_{L}e^{-\left(2-{a}\right)\rho}+\frac{{a}}{8\left(1-{a}\right)}\newg^{2}
x_{R}^{-4\frac{1-{a}}{2-{a}}}e^{-2\left(1-{a}\right)\rho}\right].\label{bc01c}
\end{eqnarray}
One can also compute the conserved quantity (\ref{ab08})
\begin{equation}
\mathcal{C}=-2x_R x_L \,.\label{CC}
\end{equation}

We are ready to formulate our asymptotic conditions. We request that all fields approach at $\rho\to\infty$ the asymptotic expressions (\ref{bc01a}), (\ref{bc02}), (\ref{bc01b}), and (\ref{bc01c}). In these formulas, $\mathcal{C}$ has to be understood as a shorthand notation for the right hand side of Eq.\ (\ref{CC}). The functions $x_R(\tau)$, $x_L(\tau)$, $\mathcal{L}(\tau)$, and $\newg(\tau)$ are free boundary data.

The equations of motion of dilaton gravity which contain $\tau$-derivatives yield equations of motion for the boundary data
\begin{eqnarray}
\newg &=& \frac{2}{2-{a}}x_{R}^{-\frac{{a}}{2-{a}}}\dot{x}_{R}, \label{bc03a}\\
\dot{x}_{L} &=& -x_{R}^{-1}\dot{x}_{R}x_{L},\label{bc03b}\\
\frac{1}{2-{a}}\ddot{x}_{R} &=&\left(1-{a}\right)\mathcal{L}x_{R}+\frac{1}{\left(2-{a}\right)^{2}}\dot{x}_{R}^{2}x_{R}^{-1}. \label{bc03c}
\end{eqnarray}
We stress, that these equations are not a part of our asymptotic conditions. They rather describe classical dynamics in the holographic theory. Eq.\ (\ref{bc03b}) is equivalent to the condition $\partial_\tau\, \mathcal{C}=0$. The meaning of (\ref{bc03c}) will be discussed in section \ref{sec:Sch}.

For a static configuration, we can match the asymptotic solutions obtained here to the exact one, see Appendix \ref{sec:ex}. Let us compare the term with $\dd \theta^2$ in the exact metric to the term with $\dd\tau^2$ in the asymptotic one in the limit $\Phi\to\infty$. For the exact metric, we have 
\begin{equation}
    \dd s^2=\xi e^Q\dd\theta^2 +\dots \simeq \Phi^{2-2a}\dd\theta^2 +\dots\simeq \tilde\Phi^{\frac{2(2-2a)}{2-a}}\dd\theta^2+\dots\simeq x_R^{\frac{2(2-2a)}{2-a}}e^{2(1-a)\rho}\dd\theta^2+\dots,
\end{equation}
where we used (\ref{S05}), (\ref{eQw}), (\ref{ab07}), and (\ref{bc01a}). In the same limit, the asymptotic line element (\ref{ab10}) reads $\dd s^2 =h^2\dd\tau^2+\dots\simeq e^{2(1-a)\rho}\dd\tau^2+\dots$. These two line elements match if $\dd\tau=x_R^{\frac{(2-2a)}{2-a}}\dd\theta$ which yields the following expression for the period of $\tau$:
\begin{equation}
\beta_\tau = x_R^{\frac {2(1-a)}{2-a}}\beta_\theta \label{betatau}
\end{equation}
with $\beta_\theta$ defined in (\ref{betat}). We stress, that (\ref{betatau}) has been derived for static configurations when $x_R$ is a constant.

The asymptotic conditions derived above can be considered as generalizations of the Fefferman-Graham expansion which are actually double expansions in powers of both $e^\rho$ and $e^{a\rho}$. 

In the limit $a\to 0$ the subleading powers of $e^\rho$ in the asymptotic conditions merge leading to rearrangements of the space of boundary data.

\section{Asymptotic symmetry algebra}\label{sec:asa}
The asymptotic symmetry algebra is formed by the bulk gauge generators which map solutions of certain asymptotic conditions to themselves. This means, that this algebra is realized on the boundary data $x_R(\tau)$, $x_L(\tau)$, $\mathcal{L}(\tau)$, and $\newg(\tau)$ and can be obtained by finding all asymptotic Killing vectors for given asymptotic conditions and extracting their action on the boundary data. With the gauge fixed form of the asymptotic line element (\ref{ab10}) and after perturbatively solving the equations of motion in the first-order formulation leading to (\ref{bc02}) we are ready to compute the asymptotic Killing vectors $\xi$. This is done by performing a Lie-variation of the metric
\begin{equation}
    \delta _\xi g_{\mu\nu} =\xi ^\alpha \partial _\alpha g_{\mu \nu }+g_{\mu \alpha }\partial _\nu \xi ^\alpha +g_{\alpha \nu }\partial _\mu \xi ^\alpha 
    \label{AKV_def}
\end{equation}
and demanding that it does not change the form of the line element in terms of an expansion in powers of $e^\rho $. In our situation the asymptotic conditions on the line element (\ref{bc02}) suggest to fix the leading order terms and let the coefficients of all subleading orders vary.
We therefore choose
\begin{equation}
\delta g_{\rho\rho} = 0,\;\;\;\;\delta g_{\rho\tau} = 0, \;\;\;\delta g_{\tau\tau} = 2\delta \mathcal L + \delta \mathcal L^2e^{-2(1-a)\rho }+\mathcal{O}(e^{-a\rho })
\end{equation}
and readily obtain three relations from (\ref{AKV_def})
\begin{subequations}
\begin{eqnarray}
\delta _\xi g_{\rho \rho } &=&2\partial _\rho \xi ^\rho =0,\\
\delta _\xi g_{\tau \rho } &=& \partial _\tau \xi ^\rho +h^2\partial _\rho \xi ^\tau =0,\\
\delta _\xi g_{\tau \tau } &=& \xi ^\alpha \partial _\alpha h^2+2h^2\partial _\tau \xi ^\tau =\delta g_{\tau\tau} .
\end{eqnarray}
\end{subequations}
These equations now have to be solved for $\xi ^\alpha $. The first two of them yield
\begin{eqnarray}
\xi^\rho &=& \eta (\tau),\\
\xi^\tau &=& \epsilon(\tau) + \frac{\dot \eta}{2(1-a)}e^{-2(1-a)\rho} + \mathcal O( e^{-4(1-a)\rho}).
\end{eqnarray}
The solutions depend on two arbitrary functions $\epsilon$ and $\eta$. From the third equation an additional constraint follows
\begin{equation}
    \eta = -\frac{\dot \epsilon}{1-a},
\end{equation}
which reduces the number of parameters by one. The final expression for the asymptotic Killing vectors thus reads
\begin{eqnarray}
\xi^\rho &=& -\frac{\dot \epsilon}{1-a}, \label{akv1}\\
\xi^\tau &=& \epsilon - \frac{\ddot \epsilon}{2(1-a)^2}e^{-2(1-a)\rho} + \mathcal O( e^{-4(1-a)\rho}).\label{akv2}
\end{eqnarray}
As a consistency check the Lie bracket between two of these vector fields can be computed which leads to
\begin{eqnarray}
\big [\xi (\epsilon _1),\xi (\epsilon _2)\big ]_\mathrm{Lie} &=& \xi (\epsilon _1\dot{\epsilon }_2-\epsilon _2\dot{\epsilon }_1) + \mathcal O( e^{-4(1-a)\rho})
\end{eqnarray}
and shows that they indeed asymptotically span an algebra. By choosing properly normalized Fourier modes for the transformation parameters and defining 
\begin{align}
    \ell _n:=\xi \Big (\epsilon =\frac{\beta _\tau }{2\pi }e^{\frac{i2\pi n\tau }{\beta _\tau }}\Big )
\end{align}
a Witt algebra
\begin{eqnarray}
i\big [\ell _n,\ell _m\big ]_\mathrm{Lie}=(n-m)\ell _{n+m} \label{akv_bracket}
\end{eqnarray}
can be identified.
With $\Lambda := x^{-2}_R\mathcal C$ the variation of the fields in the metric is
\begin{eqnarray}
\delta_\xi \mathcal L &=& \epsilon \dot{\mathcal L} + 2\dot \epsilon \mathcal L - \frac{\dddot \epsilon}{2(1-a)^2}, \label{asaL}\\
\delta_\xi \Lambda &=& \dot{\Lambda} \epsilon + \frac{2-a}{1-a}\Lambda \dot \epsilon,\label{lambda_trafo}
\end{eqnarray}
and taking the Lie derivative of the asymptotic expression for the dilaton \eqref{bc01a} yields the transformation rules of the fields $x_R$, $x_L$ and $\ell$
\begin{eqnarray}
\delta_\xi x_R &=& \epsilon \dot{x}_R  - \frac{2-a}{2(1-a)}\dot\epsilon x_R ,\label{asaXR}\\
\delta_\xi x_L &=& \epsilon \dot x_L + \frac{2-a}{2(1-a)}\dot\epsilon x_L ,\label{asaXL}\\
\delta_\xi \ell &=& \epsilon \dot \ell - \frac{2}{(2-a)(1-a)}\frac{\dot x_R}{\ell} x^{\frac{2-3a}{2-a}}_R \ddot \epsilon .\label{asaELL}
\end{eqnarray}
These infinitesimal transformations have to preserve the equations of motion governing the boundary dynamics (\ref{bc03a})-(\ref{bc03c}) which can be checked to hold here by computing their variation and evaluating it on-shell.  
As there is only one free function in the asymptotic Killing vectors, there will only be one tower of charges generating the above transformations. From \eqref{asaL} it can be seen that $\mathcal L$ transforms with an infinitesimal Schwarzian derivative and thus behaves like a CFT stress tensor. By observing the transformation behaviour of $x_R$ in (\ref{asaXR}) we can define  
\begin{eqnarray}
 y\equiv x_{R}^{\frac{2\left(1-a\right)}{2-a}} \label{defy}
\end{eqnarray}
which leads to
\begin{equation}
\delta_\varepsilon y=y\dot{\varepsilon}-\varepsilon\dot{y} \label{gty}
\end{equation}
telling us that $y$ transforms as a boundary vector. This last equation can be written as $\delta_\varepsilon (1/y)=-\partial_\tau (\varepsilon /y)$. Thus, if we define the average value 
\begin{equation}
\langle  1/y \rangle \equiv \frac 1{\beta_\tau} \int_0^{\beta_\tau} \dd\tau \, \frac 1y \label{y_mean_def}
\end{equation}
we have
\begin{equation}
\delta_\varepsilon \langle 1/y \rangle =0 \,.\label{de1y}
\end{equation}
This equation will be useful in section \ref{sec:boa}.

At this point, it would be natural to construct the canonical asymptotic charges which generate the symmetries obtained above and are finite and integrable with our asymptotic conditions. This is another example of a task which becomes very nontrivial if one considers fairly generic families of dilaton gravities. A complete analysis of the asymptotic charges for models with $U=0$ was performed recently in \cite{Ruzziconi:2020wrb}, but this class does not include our models. Fortunately, we can bypass this step if want to identify the symmetry algebra only.

Since Eq.\ (\ref{asaL}) is nothing else than the coadjoint action of the Virasoro algebra, we conclude that the Virasoro generators $\mathcal{L}_n$ have to be the Fourier model of $\mathcal{L}$,
\begin{equation}
    \mathcal{L}(\tau)=\gamma \sum_{n\in\mathbb{Z}}\mathcal{L}_n e^{-\frac{i2\pi \tau}{\beta_\tau}}\label{Ln}
\end{equation}
up to an undetermined parameter $\gamma$. On the same grounds, the charge which generates this transformation should be
\begin{equation}
    Q(\epsilon) =\frac{\alpha}{\beta_\tau}\int_0^{\beta_\tau} \dd\tau  \epsilon(\tau) \mathcal {L}(\tau) \label{charge}
\end{equation}
with another undetermined parameter $\alpha$. One has to keep in mind that (\ref{charge}) uses the trick of defining averaged charges which was previously also applied in \cite{Grumiller:2017qao,Cadoni:2000ah}. 
The transformation (\ref{asaL}) has to be given by a Lie algebra bracket $\delta_\epsilon\mathcal{L}(\tau)=[Q(\epsilon),\mathcal{L}(\tau)]$. This is consistent with the Virasoro commutation relations
\begin{equation}
i[\mathcal L_n , \mathcal L_m] = (n-m) \mathcal L_{m+n} + \frac{c}{12} \delta_{m+n,0}
\label{vir_alg}
\end{equation}
if the constants satisfy the following conditions:
\begin{equation}
    \alpha=-\frac{2\pi}{\beta_\tau \gamma},\qquad \gamma=\frac{24\pi^2}{c\beta_\tau^2 (1-a)^2 }.
\end{equation}
This result is in correspondence with the charge representation theorem stating that up to central extensions the canonical charges fulfil the same algebra as the asymptotic Killing vectors. In this case, the other boundary fields are related to pure gauge transformations. It should be noted that the central charge remains an undetermined constant, the only information we can get from the above is that it is non-zero. We want to point out that the more standard representation of Virasoro commutation relations
\begin{align}
    i[\bar{\mathcal L}_n , \bar{\mathcal L}_m] = (n-m) \bar{\mathcal L}_{m+n} + \frac{c}{12}(n^3-n) \delta_{m+n,0}
\end{align}
can be reached by shifting the zero mode $\mathcal{L}_n = \bar{\mathcal{L}}_n-\frac{c}{24}\delta _{n,0}$.

Although  we cannot fix the value of central charge by our arguments, it is still interesting to have a look at other aspects on the CFT side. For example one might expect that for these given Virasoro symmetries the Cardy formula reproduces the right result for black hole entropy, analogous to the JT-case \cite{Grumiller:2017qao}. However, it will be seen in section \ref{sec:entropy} that this is not the case. Instead we find that a generalized Cardy formula might be applicable.

\section{Boundary action}\label{sec:boa}
To obtain a consistent theory one needs to add certain boundary terms to the bulk action. Such terms serve two purposes. They make the variational problem consistent and the on-shell action finite. In this section, we suggest a boundary action and show that it is compatible with our asymptotic conditions.

The necessary boundary terms are easier to find and analyse in the second-order formulation. We propose the following action 
\begin{equation}\label{v01}
\Gamma=I_{\mathrm{2nd}}+I_{\mathrm{GHY}}+I_{\mathrm{CT}}+I_{\mathrm{K}}
\end{equation}
with
\begin{eqnarray}
I_{\mathrm{GHY}} &=&	-\int\mathrm{d}\tau\;\sqrt{\gamma}\Phi K, \label{v02a}\\
I_{\mathrm{CT}}	&=&	\int\mathrm{d}\tau\sqrt{\gamma} \Phi , \label{v02b}\\
I_{\mathrm{K}} &=& \frac{1}{2}\int\mathrm{d}\tau\;\sqrt{\gamma}\Phi^{-1}\gamma^{\tau\tau}\partial_{\tau}\Phi
\partial_{\tau}\Phi \,. \label{v02c}
\end{eqnarray}
Here $\gamma$ is the induced metric and $K$ is the trace of the extrinsic curvature of the asymptotic boundary. $I_{\mathrm{GHY}}$ is the usual Gibbons-Hawking-York term. It is accompanied by two holographic counterterms. The action $\Gamma$ coincides with the one proposed in \cite{Grumiller:2017qao} for the JT model for some special choice of a free constant in that action. It is somewhat surprising that this action also works for more general models even though it is known \cite{Grumiller:2007ju} that this action is consistent for a dilaton which does not fluctuate at leading order, $x_R=const.$

First, we check the consistency of the variational principle. The general outline of this procedure is borrowed from \cite{Grumiller:2017qao}. We vary $\Phi$ and the metric in $\Gamma$, eq.\ (\ref{v01}), assuming only that the fields satisfy our asymptotic conditions. Afterwards, we impose the equations of motion. All bulk contributions vanish automatically, while the boundary terms read
\begin{eqnarray}\label{v06}
(\delta\Gamma)_{\mathrm{on-shell}}=-\int_{\partial\mathcal{M}}\mathrm{d}\tau\; (Q_{h}\delta h+ Q_{\Phi}\delta\Phi)
\end{eqnarray}
where
\begin{eqnarray}
Q_{h} &=& \partial_{\rho}\Phi-\Phi+\frac{1}{2}\Phi^{-1}h^{-2}\left(\partial_{\tau}\Phi\right)^{2} \label{v07a}\\
Q_{\Phi} &=& {a}h\Phi^{-1}\partial_{\rho}\Phi+\partial_{\rho}h-h+\frac{1}{2}h^{-1}\left(\Phi^{-1}\partial_{\tau}
\Phi\right)^{2}+\partial_{\tau}\left(h^{-1}\Phi^{-1}\partial_{\tau}\Phi\right). \label{v07b}
\end{eqnarray}
We now use the asymptotic conditions to relate $\delta \Phi$ and $\delta h$ to $\delta x_R$, $\delta x_L$, $\delta \mathcal{L}$, and $\delta \newg$. The equations of motion \eqref{bc03a} \eqref{bc03b}, \eqref{bc03c} on $x_R$, $x_L$, $\mathcal{L}$, and $\newg$ are used to simplify the expressions. After long but straightforward calculations, we obtain that all terms proportional to $\delta x_L$, $\delta \mathcal{L}$, and $\delta \newg$ cancel out. The remaining term reads
\begin{equation}
(\delta\Gamma)_{\mathrm{on-shell}}=-\int_{\partial\mathcal{M}}\mathrm{d}\tau\;\mathcal{C}\delta (y^{-1}). \label{varG}
\end{equation}
Since $\mathcal{C}$ is constant on-shell the variation (\ref{varG}) vanishes if the constant mode of $y^{-1}$ does not vary,
\begin{equation}
\delta \langle 1/y \rangle =0.\label{varcony}
\end{equation}
We shall treat (\ref{varcony}) as an additional boundary condition. Note, that due to (\ref{de1y}) this condition does not reduce the asymptotic symmetry algebra.

Let us now compute the on-shell value of $\Gamma$. We already know that the first order action vanishes on shell for our family of dilaton gravities, see (\ref{I0}). Therefore, the second order action is just the difference term (\ref{I21}), $I_{\rm 2nd}=-\int_{\partial\mathcal{M}}\Phi\, t$.  By substituting the asymptotic conditions  we obtain
\begin{eqnarray}
&&I_{\mathrm{2nd}}	=	 \int\mathrm{d}\tau\;\left[-{a}x_{R}^{\frac{2}{2-{a}}}e^{\left(2-{a}\right)\rho}+
\left(-{a}\mathcal{L}x_{R}^{\frac{2}{2-{a}}}+\frac{\left(1-2{a}\right){a}}
{\left(1-{a}\right)\left(2-{a}\right)^{2}}x_{R}^{-\frac{2\left(1-{a}\right)}{2-{a}}}
\dot{x}_{R}^{2}\right)e^{{a}\rho}\right. \nonumber\\
&&\qquad\qquad\qquad\qquad\qquad\qquad\qquad\qquad \left. -{a}\mathcal{C}x_{R}^{-\frac{2\left(1-{a}\right)}
{2-{a}}}\right] . \label{VA05a}
\end{eqnarray}
Similarly, we compute the actions (\ref{v02a}) - (\ref{v02c})
\begin{eqnarray}
&&I_{\mathrm{GHY}}	=	 \int\mathrm{d}\tau\;\left[-\left(1-{a}\right)x_{R}^{\frac{2}{2-{a}}}e^{\left(2-{a}\right)\rho}+
\left(\left(1-{a}\right)\mathcal{L}x_{R}^{\frac{2}{2-{a}}}-\frac{1}{\left(2-{a}\right)^{2}}
x_{R}^{-\frac{2\left(1-{a}\right)}{2-{a}}}\dot{x}_{R}^{2}\right)e^{{a}\rho}
+\mathcal{C}
x_{R}^{-\frac{2\left(1-{a}\right)}{2-{a}}}\right]   \nonumber\\
&&I_{\mathrm{CT}}	=	 \int\mathrm{d}\tau\;\left[x_{R}^{\frac{2}{2-{a}}}e^{\left(2-{a}\right)\rho}+\left(x_{R}^{\frac{2}
{2-{a}}}\mathcal{L}+\frac{1}{\left(1-{a}\right)\left(2-{a}\right)^{2}}
x_{R}^{-\frac{2\left(1-{a}\right)}{2-{a}}}\dot{x}_{R}^{2}\right)e^{{a}\rho}\right]  \nonumber\\
&&I_{\rm K}=\frac{2}{\left(2-{a}\right)^{2}}\int\mathrm{d}\tau\; x_{R}^{-\frac{2\left(1-{a}\right)}{2-{a}}}\dot{x}_{R}^{2}e^{{a}\rho}.\nonumber
\end{eqnarray}
The most divergent contributions proportional to $e^{(2-a)\rho}$ cancel against each other. All other terms combine to
\begin{equation}\label{VA06}
\Gamma=\int\mathrm{d}\tau\;2x_{R}^{\frac{{a}}{2-{a}}}\left[\left(1-{a}\right)\mathcal{L}
x_{R}+\frac{\left({a}+1\right)}{\left(2-{a}\right)^{2}}x_{R}^{-1}\dot{x}_{R}^{2}\right]
e^{{a}\rho}+\int\mathrm{d}\tau\;\left(1-{a}\right)\mathcal{C}x_{R}^{-\frac{2\left(1-{a}\right)}
{2-{a}}}.
\end{equation}
Here, we use the equation of motion (\ref{bc03c}) and integrate by parts to see that the term with $e^{a\rho}$ also goes away. Thus, we are left with a finite expression
\begin{equation}\label{v03}
\Gamma \big \vert _{\mathrm{on-shell}}= \int_{\partial\mathcal{M}}\mathrm{d}\tau\;\left(1-{a}\right)\mathcal{C}x_{R}^{-\frac{2\left(1-{a}\right)}{2-{a}}}
=\left(1-{a}\right)\int\mathrm{d}\tau\;\mathcal{C}y^{-1}.
\end{equation}

If the condition $a<2/3$ is fulfilled the terms written explicitly in the expansions (\ref{bc01a}) and (\ref{bc02}) are sufficient to recover all terms in $\Gamma$ and $\delta\Gamma$ which are divergent or finite in the limit $\rho\to\infty$.

\section{Entropy formulae}\label{sec:entropy}
On the gravity side, we have several possibilities available to check the consistency of our action principle regarding thermodynamic properties. One important aspect is black hole entropy. For this it is assumed that we have a static black hole solution which implies $\mathcal{L} =0$ by the equations of motion (\ref{bc03a})-(\ref{bc03c}). As discussed in appendix \ref{sec:ex} the Casimir $\mathcal C$ has to be a negative constant in order to avoid a naked singularity. There it is also shown that regularity at the Euclidean horizon yields a certain inverse temperature (\ref{betat}) which for static solutions can be adapted to the main part by (\ref{betatau}). With the redefinition (\ref{defy}) this yields 
\begin{align}
    \beta _\tau =\frac{1}{y}\frac{4\pi }{(2-a)\Phi _0^{1-a}}=\Big \langle \frac{1}{y} \Big \rangle \frac{4\pi }{(2-a)\Phi _0^{1-a}}, \label{eucl_temp}
\end{align}
where in the last equality it was used that we have $1/y=\langle 1/y \rangle $ (cf. (\ref{y_mean_def})).

Wald's method gives a universal result for all dilaton gravity models \cite{Gegenberg:1994pv,Wald:1999wa} stating that the macroscopic entropy for such black hole solutions is related to the value of the dilaton at the horizon 
\begin{equation}
    S_\mathrm{Wald}=2\pi \Phi _0 \label{wald_entropy}
\end{equation}
which as shown in appendix \ref{sec:ex} can be expressed in terms of the Casimir function in the following way
\begin{equation}
    \Phi _0=(-2\mathcal{C})^{\frac{1}{2-a}}.
    \label{dil_at_hor}
\end{equation}
Combining (\ref{eucl_temp})-(\ref{dil_at_hor}) we obtain a relation between entropy and temperature 
\begin{align}
    S_\mathrm{Wald}(T)=2\pi \Big (\frac{4\pi }{2-a}\Big )^\frac{1}{1-a}\Big \langle \frac{1}{y} \Big \rangle ^\frac{1}{a-1}T^\frac{1}{1-a}. \label{wald_entr}
\end{align}
which is compatible with the third law for our range of the parameter $a$. 

Now let's approach the entropy computation in a different way and take the Euclidean path integral in the saddle-point approximation. It is dominated by the action evaluated on a classical solution, which is taken to be a static black hole with our prescribed boundary conditions of section \ref{sec:asc}. In this approximation we can relate
\begin{align}
    \Gamma \big \vert _{\mathrm{on-shell}} =\beta _\tau F,
\end{align}
where $F$ is the free energy of the static black hole solution and $\beta _\tau =1/T$. Evaluating the on-shell action (\ref{v03}) with the definition (\ref{y_mean_def}) and using the on-shell constancy of the Casimir gives
\begin{align}
    \Gamma \big \vert _{\mathrm{on-shell}}=(1-a)\beta _\tau \mathcal{C} \Big \langle \frac{1}{y}\Big \rangle
\end{align}
which can be used to express $F$ in terms of $\mathcal{C}$. We furthermore know that for such a thermal state the entropy is related to the free energy by
\begin{align}
    S(T)=-\frac{\partial F}{\partial T} =-(1-a)\Big \langle \frac{1}{y}\Big \rangle\frac{\partial }{\partial T}\mathcal{C}(T) \label{entr_freeE}
\end{align}
where the Casimir as a function of the temperature is determined by (\ref{eucl_temp}) and (\ref{dil_at_hor}). Inserting
\begin{align}
    \mathcal{C}(T)=-\frac{1}{2}\Big (\frac{4\pi }{2-a}\Big )^\frac{2-a}{1-a}\Big \langle \frac{1}{y}\Big \rangle ^\frac{2-a}{a-1}T^\frac{2-a}{1-a}. \label{casimir_T}
\end{align}
into (\ref{entr_freeE}) precisely reproduces the result (\ref{wald_entr}). 

We can also check the first law $TdS=d\Delta $ with the state dependent energy $\Delta $ and find that it indeed holds provided we set
\begin{align}
    \Delta =-\Big \langle \frac{1}{y}\Big \rangle \mathcal{C}.
\end{align}
This gives the on-shell Casimir the natural interpretation of a black hole mass for our chosen models. By using (\ref{casimir_T}) we arrive at a generalized Stefan--Boltzmann law 
\begin{align}
    \Delta =\frac{1}{2}\Big (\frac{4\pi }{2-a}\Big )^\frac{2-a}{1-a}\Big \langle \frac{1}{y}\Big \rangle ^\frac{1}{a-1}T^\frac{2-a}{1-a}.\label{energy}
\end{align}

Having had a look at the gravity side we switch to the field theory side. For this we might start from the Virasoro asymptotic symmetry algebra (\ref{vir_alg}) and give in to the temptation of using the Cardy formula to compute the entropy. For the JT-model this indeed turned out to work \cite{Grumiller:2017qao} yielding the same result for black hole entropy as the computation on the gravity side. It shall now be demonstrated that it however is inconsistent for our class of models with $a\neq 0$ as the central charge would have to acquire state dependence. Starting from the Virasoro algebra given in terms of the modes $\mathcal{L}_n$ by (\ref{vir_alg}) we insert into the Cardy formula\footnote{Note that we have to work with the version of the Cardy formula which has a shifted zero mode as the charges $\mathcal{L} _n$ obey the non-standard bracket relation (\ref{vir_alg}).} 
\begin{align}
    S=2\pi \sqrt{\frac{c}{6}\Big (\mathcal{L}_0+\frac{c}{24}\Big )}. \label{cardy_formula}
\end{align}
It is clear that as in our context the central charge is unknown a direct application of (\ref{cardy_formula}) is impossible. We could however approach from the other end and assume for a moment that the Cardy formula holds. Taking again a static black hole solution it might be possible to determine the central charge by matching expression (\ref{wald_entropy}) with the LHS of (\ref{cardy_formula}). As for such a solution (\ref{bc03c}) implies $\mathcal{L} _0=0$ we would directly get a result of the form
\begin{align}
    c=c(\mathcal{C})
\end{align}
which is a state dependent central charge and makes the Cardy formula inapplicable.

Being not consistent with the Cardy formula does not mean that other CFT entropy formulas are also ruled out. Let us consider an example of entropy formula which does work. This formula has been suggested in the context of three dimensional gravity \cite{Perez:2016vqo,Afshar:2016kjj,Gonzalez:2011nz} in which the symmetries of the boundary metric involve anisotropic scalings along the two boundary coordinate directions. For such theories with anisotropic Lifshitz-scaling it was shown that a generalized Cardy formula could reproduce the correct result for black hole entropy. In our case we only have a one-dimensional boundary manifold which makes it somewhat delicate to talk about anisotropy in the first place. We however still want to state a possible two-dimensional version of the formula presented in the above references:
\begin{equation}
    S=2\pi (1+z)\Big [\Delta \Big (\frac{\Delta _0[1/z]}{z}\Big )^z\Big ]^\frac{1}{1+z}.
    \label{anis_cardy}
\end{equation}
The quantity $\Delta $ is the energy of the state (black hole) whose entropy we compute and $-\Delta _0[z]$ is the ground state energy. In some sense it plays the role of the central charge for theories with anisotropic scaling. The parameter $z$ describes the amount of anisotropy with the value $z=1$ corresponding to the isotropic case in which (\ref{anis_cardy}) reduces to (\ref{cardy_formula}) provided we set $\Delta =\mathcal{L}_0+\frac{c}{24}$ and $\Delta _0=\frac{c}{24}$. We observe, that provided we identify the state dependent energy $\Delta $ with (\ref{energy}) and set the anisotropy parameter to 
\begin{align}
    z=1-a
\end{align}
we get the exact same temperature dependence as in (\ref{wald_entr}) , $S\propto T^{1/(1-a)}$. Also, the isotropic case precisely corresponds to the JT-model (a=0) in which indeed the standard Cardy formula holds. 

Of course, this is by far not sufficient evidence for the applicability of (\ref{anis_cardy}) but it still suggests that the field theory side of the class of models described here is not of the same structure as in JT-gravity. It might be the case that the formula incidentally leads to a matching entropy but the parameter $z$ has to be interpreted in a different way than in the three-dimensional case. Also, we cannot make any prediction on how to obtain the ground state energy $-\Delta _0$. This is mainly due to the missing link between certain Virasoro coadjoint orbits and the corresponding spacetime geometries. If just a link could be found, the ground state energy could be brought into correspondence with a maximally symmetric spacetime solution like this was done in \cite{Afshar:2016kjj}.

\section{Boundary dynamics and the Schwarzian action}\label{sec:Sch}

The asymptotic conditions depend on four functions $x_L$, $x_R$, $\newg$, and $\mathcal{L}$ of the boundary coordinate $\tau$. Due to (\ref{bc03a}), the function $\newg(\tau)$ is uniquely expressed through $x_R(\tau)$. By virtue of (\ref{CC}), $x_L(\tau)$ depends on $x_R(\tau)$ and the integration constant $\mathcal{C}$. Thus, the essential variables describing the classical asymptotic dynamics are $x_R$ and $\mathcal{L}$, besides the constant $\mathcal{C}$.

The transformation properties of $\mathcal{L}$, see section \ref{sec:asa}, show that this field marks the points on co-adjoint orbits of the Virasoro group.

To clarify the meaning of the remaining degree of freedom $x_R$ let us rewrite equation (\ref{bc03c}) in terms of the related variable $y$, eq.\ (\ref{defy}),
\begin{equation}
0=-4(1-a)^2\mathcal{L} + \frac{\dot{y}^2}{y^2} +2\partial_\tau \left( \frac{\dot{y}}y \right). \label{yLeq}
\end{equation}
We multiply both sides of this equation with $y^2$ and take the derivative with respect to $\tau$ to obtain
\begin{equation}
0=-2(1-a)^2(\dot{\mathcal{L}}y+2\mathcal{L}\dot{y})+\dddot{y}.\label{0011}
\end{equation}
This should be compared with the asymptotic transformation law for $\mathcal{L}$, eq.\ (\ref{asaL}). Thus, eq.\ (\ref{0011}) is equivalent to
\begin{equation}
\delta_y \left( 2(1-a)^2\mathcal{L}\right)=0,
\end{equation}
which states that $y$ is a stabilizer of $\mathcal{L}$.

The equation (\ref{yLeq}) is generated by the action 
\begin{equation}
S=\int_0^{\beta_\tau}\dd\tau\, \left[ 4(1-a)^2y\mathcal{L} + \frac{\dot{y}^2}y-2\ddot{y}\right] \label{SyL1}
\end{equation}
upon variation with respect to $y$. We added to the action a total derivative term which will be useful later. We stress that here $\mathcal{L}$ plays the role of an external parameter and is not varied. Let us define a function $v(\tau)$ such that
\begin{equation}
\dot{v}= \frac 1{y\langle 1/y\rangle}.\label{utau}
\end{equation}
From this definition, $v$ has to be quasi periodic, $v(\tau +\beta_\tau)=v(\tau)+\beta_\tau$. In terms of $v$, the action (\ref{SyL1}) reads
\begin{equation}
S=\frac 1{\langle 1/y \rangle}\int_0^{\beta_\tau}\dd\tau\,\frac 1{\dot{v}} \left[ 4(1-a)^2\mathcal{L}+2\, \mathrm{Sch}[v](\tau) \right],\label{SuL}
\end{equation}
where $\mathrm{Sch}[v](\tau)$ denotes the Schwarzian derivative,
\begin{equation}
\mathrm{Sch}[v](\tau)\equiv \frac{\dddot{v}}{\dot{v}} - \frac 32 \left( \frac {\ddot{v}}{\dot{v}} \right)^2 .\label{Sch}
\end{equation}
By changing the integration variable $\tau\to v$ and using the inversion formula for Schwarzian derivative we arrive at the action
\begin{equation}
S=\frac 1{\langle 1/y \rangle}\int_0^{\beta_\tau} \dd v \left[ 4(1-a)^2\mathcal{L}\, (\partial_v\tau)^2 - 2\, \mathrm{Sch}[\tau](v) \right].\label{SLtau}
\end{equation}
Up to setting $\mathcal{L}$ to a suitable constant, this is the Schwarzian action proposed in \cite{Maldacena:2016hyu,Maldacena:2016upp}. 

In the paper \cite{Grumiller:2017qao} the Schwarzian action for JT model was obtained by going "partially off-shell" in the on-shell action (\ref{v03}). For $a\neq 0$ this method does not work since relaxing the equations of motions immediately leads to a divergence in $\Gamma$ at $\rho\to\infty$, cf (\ref{VA06}). We saw that a Schwarzian boundary action can only be obtained by the equivalence of its equations of motion to a boundary equation of motion. The true boundary theory might therefore contain something more than just a Schwarzian action.  

\section{Conformally related theories}
\label{sec:con}
Having obtained a consistent set of asymptotic conditions and a boundary action for a family of dilaton gravities we can extend this to an even larger family of models by making a dilaton-dependent conformal transformation of the metric.

Let us consider a dilaton gravity model with a metric $\hat g_{\mu\nu}$, a dilaton field $\Phi$, and dilaton potentials $\hat U(\Phi)$ and $\hat V(\Phi)$. Let us make a dilaton-dependent Weyl rescaling of the metric, 
\begin{equation}
\hat g_{\mu\nu}=e^{2\rho (\Phi)} g_{\mu\nu}\,.\label{hatg}
\end{equation}
In terms of the new metric one obtains again a dilaton gravity with the potentials
\begin{equation}
    U(\Phi)=\hat U(\Phi) -2 \frac {\dd}{\dd\Phi}\, \rho(\Phi), \qquad V(\Phi)=e^{2\rho(\Phi)}\hat V(\Phi).\label{hatUV}
\end{equation}
These two dilaton gravities look formally equivalent but their physical contents are in fact different. This difference becomes apparent if one couples both theories to the same matter theory. Minimally coupled point particles should follow geodesic lines -- but of two different metrics. The rescaling (\ref{hatg}) can create or remove horizons and curvature singularities and change the global structure of space-time. For quantum matter fields, this rescaling leads to a conformal anomaly. 

Suppose that we have a consistent boundary value problem and an interesting holographic dual for the dilaton model with $\hat g$, $\hat U$, and $\hat V$. By simply rewriting the action together with boundary terms and the asymptotic conditions in terms of $g$, $U$, and $V$ we obtain a consistent boundary value problem and the same holographic dual for the conformally related model. This simple procedure generates holographic correspondences which become nonequivalent upon inclusion of matter fields. We are not going to include the matter explicitly. For the moment, it is sufficient to recognize that different metric geometries lead to different physical realities.  

The model with $\hat g$, $\hat U$ and $\hat V$ has to belong to the family (\ref{UV}) since these are the models for which we know the asymptotic conditions and the boundary action. After the rescaling (\ref{hatg}), the total boundary action becomes 
\begin{eqnarray}
  I_{\mathrm b}&=&  \int_{\partial\mathcal M}\dd\tau\, \sqrt{\gamma}\left( - \Phi K + e^{\rho(\Phi)}\Phi + \frac{1}{2} e^{-\rho(\Phi)} \Phi^{-1} \gamma^{\tau\tau}(\partial_\tau \Phi)^2 \right) \nonumber\\
  &=& \int_{\partial\mathcal M}\dd\tau\, \sqrt{\gamma}\left( - \Phi K + \sqrt{ -2w(\Phi)e^{-Q(\Phi)}} +\frac{\gamma^{\tau\tau}(\partial_\tau \Phi)^2}{2\sqrt{ -2w(\Phi)e^{-Q(\Phi)}}}\right)
  .\label{Ib}
\end{eqnarray}
Modulo a bit different notations and conventions this action coincides with the boundary action used in \cite{Grumiller:2017qao} for the JT model and with the boundary action proposed in \cite{Grumiller:2007ju} for generic dilaton gravities with a dilaton field which does not fluctuate at the asymptotics ($\partial_\tau\Phi=0$). We may even conjecture that the same boundary action (\ref{Ib}) is valid for all dilaton gravity models with a fluctuating dilaton for a suitable choice of asymptotic conditions. Note, however, that at the moment we do not even have a good guess for asymptotic conditions valid for any dilaton gravity. 

The are no a priori restrictions on the the form $\rho$. To illustrate the method, let us take JT as the untransformed model and require that the transformed models remains inside the $(a,b)$ family (\ref{ab01}), i.e., that $U=-(a/\Phi)$ for some $a$. Then,
\begin{equation}
    \rho=\frac a2\, \ln\Phi,\qquad V=-\frac B2\, \Phi^{a+1}.\label{UVconf}
\end{equation}
This means that by these transformations we may reach any point on the $b=1$ line within the $(a,b)$ plane.\footnote{One can easily check that starting with other allowed models on the $(a,b)$ plane by using conformal rescalings one can move along the $b=const$ lines.} The curvature scalar (\ref{Rab}) for these models reads
\begin{equation}
    R=-2a\mathcal{C}\Phi^{a-2} -(2-a)\Phi^a .\label{Ra}
\end{equation}
Here we took $B=2$ which is the usual choice for JT model. The behaviour of $R$ is defined by competition of two different powers of $\Phi$ in (\ref{Ra}). There are two asymptotic regions, $\Phi\to\infty$ and $\Phi\to 0$, where these powers have to be compared. From the great variety of possible cases one can select some typical situations:

\begin{enumerate}
    \item $a<0$. At $\Phi\to \infty$, $R\to 0$. Thus, we deal with flat spaces in this asymptotic region. The first term in (\ref{Ra}) decays faster than the second term. We have an example of asymptotically flat space holography. The holographic theory differs significantly from the other example of 2D flat holography \cite{Afshar:2019axx} where a twisted warped conformal theory was obtained (which is a specific limit of the complex SYK model). At $\Phi\to\infty$ the line element reads
    \begin{equation}
        \dd s^2\simeq \dd \zeta ^2 + \left( -\frac a2 \zeta \right)^{-\frac{2(2-a)}a}\dd\theta^2,\label{ds2u}
    \end{equation}
    where $\zeta =-\frac 2a \Phi^{-\frac a2}$. Although this metric is asymptotically Ricci flat at $\zeta \to\infty$, it does not describe a flat Euclidean finite temperature geometry with respect to the coordinate $\theta$ which is cyclic at the horizon. We believe that the right name for such geometries is "asymptotically trumpet".
    \item $a\in (0,2)$. Both regions $\Phi\to 0$ and $\Phi\to\infty$ correspond to curvature singularities. In the former case, the strength of the singularity is defined by the second term in (\ref{Ra}), while in the latter one by the first term which depends on the integration constant $\mathcal{C}$. For the JT model, the asymptotic region is at $\Phi\to\infty$. This is the only region which can be treated by the conformal transformation method. We have a holographic correspondence at the curvature singularity.
\end{enumerate}

The conformal transformations described above are smooth and invertible at the horizons (tips of the Euclidean metric). Therefore, they do not change the value of dilaton at the horizon (which defines the entropy) and the period of Euclidean time of the near horizon metric (which defines the temperature). For this reason, the thermodynamics properties of black holes remain intact. However, physical interpretation of these geometries remains unclear. We thus refrain from going into more details.

\section{A different gauge and some more $(a,b)$-models}\label{sec:more}
In this section, we study the asymptotic symmetry algebras for more loose asymptotic conditions and more dilaton gravities on the $(a,b)$ plane. Restricting ourselves to the asymptotic symmetries only, we thus avoid the necessity to analyse the subleading terms in asymptotic expansions for the metric and the dilaton. This latter step is necessary for checking the consistency of boundary value problem and boundary action and is the most technically demanding part of our approach.

As opposed to the Fefferman--Graham-like gauge used for the asymptotic region in the past sections it can be illuminating to choose a different one. Such an example shall be provided here by switching to Eddington--Finkelstein gauge.\footnote{We like to stress again that in the asymptotic region the gauge symmetries are (partially) broken. Thus different gauges lead to physically inequivalent asymptotic theories.} 
Taking the exact solution (\ref{S05}) for the potentials (\ref{ab01}) as a starting point it is possible to transform the line element into the form
\begin{align}
    \dd s^2=\xi e^Q \dd \theta ^2+\frac{\dd r^2}{\xi e^Q}
\end{align}
by defining a new radial coordinate such that
\begin{equation}
    \dd\Phi =\frac{\dd r}{e^Q} \quad \rightarrow \quad \Phi (r)=\big [(1-a)r]^{\frac{1}{1-a}} \label{phi_r}.
\end{equation}
After a transition to retarded time 
\begin{equation}\label{EF_trafo}
    u := \theta - ir^*, \qquad r^*=\int \frac{\dd r}{e^Q}
\end{equation}
we can recast the expression in the desired form
\begin{equation}\label{ab_lineelem}
    \dd s^2=2i\dd u \dd r+e^Q\xi \dd u^2, \qquad e^Q\xi = \Phi ^{-a}\Big [2\mathcal{C} + \frac{B}{(b+1)}\Phi ^{b+1}\Big ],
\end{equation}
where the whole model dependence is captured by the Killing norm $e^Q\xi $. The coordinate $u$ inherits its periodicity from $\theta $ by the first equation in (\ref{EF_trafo}), i.e. $\beta _u=\beta _\theta $. In the following it will be seen that starting from this gauge different symmetry algebras can arise than in the case of Fefferman-Graham gauge.

Like before the asymptotic region will be defined by a diverging dilaton which makes it possible to find the coordinate range for $r$ in dependence on the parameters $(a,b)$. The boundary conditions in each case are set to preserve the gauge 
\begin{eqnarray}
\delta g_{rr} &=&0, \\
\delta g_{ur} &=&0,
\end{eqnarray}
which according to (\ref{AKV_def}) fixes the asymptotic Killing vectors to
\begin{equation}\label{AKVs}
    \xi = \epsilon (u) \partial _u + \big (\eta (u) -\epsilon '(u)r \big )\partial _r
\end{equation}
provided $\delta g_{uu} $ and $\delta \Phi $ are not restricted too much. They are parametrized by two free functions $\epsilon (u)$ and $\eta (u)$ where $'$ denotes the derivative with respect to $u$. For a consistency check we can again compute the Lie bracket between two AKV's yielding
\begin{align}
    \big[\xi (\epsilon _1,\eta _1),\xi (\epsilon _2,\eta _2)\big] _\mathrm{Lie}=\xi \big(\epsilon _1\epsilon _2'-\epsilon _2\epsilon _1',(\epsilon _1\eta _2-\epsilon _2\eta _1)'\big),
\end{align}
which shows that they form a closed algebra. Setting $\eta =\sigma '$ and choosing Fourier modes
\begin{align}
    L_n&:=\xi \Big (\epsilon =\frac{\beta _u}{2\pi }e^{\frac{i2\pi nu}{\beta _u}},\eta =0\Big ),\\
    J_n&:=\xi \Big (\epsilon =0,\sigma =e^{\frac{i2\pi nu}{\beta _u}}\Big )
\end{align}
leads to the relations
\begin{subequations}
\begin{align}
    i\left[L_n,L_m\right]_\mathrm{Lie}&=(n-m)L_{n+m}, \label{akv_mode1}\\
    i\left[L_n,J_m\right]_\mathrm{Lie}&=-mJ_{n+m}, \label{akv_mode2}\\
    \left[J_n,J_m\right]_\mathrm{Lie}&=0. \label{akv_mode3}
\end{align}
\end{subequations}
It can be shown that the finite diffeomorphisms generated by the above asymptotic Killing vectors (\ref{AKVs}) are
\begin{equation}\label{finite_trafo}
    r \rightarrow \frac{r+g'(u)}{f'(u)}, \qquad    u \rightarrow f(u),
\end{equation}
where the functions $f(u)$ and $g(u)$ are generated by $\epsilon (u)$ and $\sigma (u)$, respectively. All of them need to be periodic with period $\beta _u$. Note that the function $g(u)$ just shifts the coordinate $r$ and thus can be interpreted as a radial supertranslation while $f(u)$ is a diffeomorphism of the circle. The group represented by this transformation is a semi-direct product 
\begin{align}
    G=\text{Diff} (S_1) \ltimes C^\infty (S_1)
\end{align}
with the group product given by
\begin{align}
    (f_1,g_1)\cdot (f_2,g_2)=(f_2\circ f_1,g_1+g_2\circ f_1).
\end{align}
It appeared before in the context of three dimensional gravity under the name warped Virasoro group, see \cite{Afshar:2015wjm} for a more extensive treatment.
The functions $g(u),f(u)$ are seen as part of the boundary field content and are used to parametrize the field space for the chosen boundary conditions.
\subsection{AdS$_2$ ground state models}
Just as in the previous sections these models are defined by $a+b=1$ which makes the Ricci scalar (\ref{Rab}) constant for $\Phi \rightarrow \infty $. For $a \in [0,\frac{3}{2})$ it can be seen from (\ref{phi_r}) that this corresponds to $r\rightarrow \infty $. After having fixed the constant $B=1+b=2-a$ the line element (\ref{ab_lineelem}) and dilaton expressed in our new coordinates take the form 
\begin{subequations}
\label{ads_metric_dil}
\begin{eqnarray}
    \dd s^2&=&2i\dd u \dd r+\Big [(1-a)^2r^2+2\mathcal{C} [(1-a)r]^{-\frac{a}{1-a}}\Big ]\dd u^2,\\
    \Phi &=&[(1-a)r]^{\frac{1}{1-a}}.\label{dil_untransformed}
\end{eqnarray}
\end{subequations}
In order to find the last part of the boundary conditions $\delta g_{uu}$ without further restricting the asymptotic Killing vectors (\ref{AKVs}) we require that  (\ref{finite_trafo}) should be allowed diffeomorphisms. By letting them act on our fields (\ref{ads_metric_dil}) we generate new orders in $r$ 
\begin{subequations}
\label{asymp_conditions}
\begin{eqnarray}
    \dd s^2 &=& 2i\dd u \dd r+\Big [(1-a)^2r^2 + P(u)r+T(u)+S(u)r^{-\frac{a}{1-a}} +\mathcal{O} (r^{-\frac{1}{1-a}})\Big ]\dd u^2,\\
    \Phi (u,r) &=& \varphi _1(u)r^{\frac{1}{1-a}}+\varphi _2(u)r^{\frac{a}{1-a}} + \mathcal{O}(r^{\frac{2a-1}{1-a}}),
\end{eqnarray}
\end{subequations}
where the redefinitions
\begin{subequations}
\begin{align}
P(u) &:= 2g'(1-a)^2-2i\frac{f''}{f'}, & \varphi _1(u) &:= \Big (\frac{1-a}{f'}\Big )^{\frac{1}{1-a}},\\
T(u) &:= (1-a)^2g'^2+2i\Big (g''-g'\frac{f''}{f'}\Big ), & \varphi _2(u) &:= \Big (\frac{1-a}{f'}\Big )^{\frac{1}{1-a}} \frac{g'}{1-a} = \varphi _1\frac{g'}{1-a},\\
S(u) &:= 2\mathcal{C}(1-a)^{\frac{a}{a-1}}f'^{\frac{2-a}{1-a}}
\end{align}\label{ads_defs}
\end{subequations}
were made. Note that only two of these functions can be allowed to vary independently - all the others are constrained by the upper relations. These constraints also could have been obtained by first choosing independent functions for each order in $r$ and then imposing the radial and dilaton equations of motion. This would be in analogy with the procedure in section \ref{sec:asc} where the equations of motion (\ref{ab14}),(\ref{ab12}) were used to fix the asymptotic conditions on the fields. By comparing the orders in $r$ the last part of the boundary conditions is chosen as 
\begin{align}
    \delta g_{uu} &= \delta P(u)r+\delta T(u)+\delta S(u)r^{-\frac{a}{1-a}} + \mathcal{O} (r^{-\frac{1}{1-a}}),\\
    \delta \Phi &= \delta \varphi _1(u)r^{\frac{1}{1-a}}+\delta \varphi _2(u)r^{\frac{a}{1-a}} + \mathcal{O} (r^{\frac{2a-1}{1-a}}),
\end{align}
so that the leading order terms in the metric are fixed.
It can be checked that this is consistent with the expression (\ref{AKVs}) for the asymptotic Killing vectors.

The boundary fields are found to transform as
\begin{eqnarray}
\delta _\xi P &=& \epsilon P'+\epsilon 'P +2(1-a)^2\eta -2i\epsilon '' ,\label{ads01}\\
\delta _\xi T &=& \epsilon T'+2\epsilon 'T+2i\eta '+P\eta, \label{ads02}\\
\delta _\xi S &=& \epsilon S'+\frac{2-a}{1-a}\epsilon 'S, \label{ads03}\\
\delta _\xi \varphi _1 &=& \varphi _1'\epsilon -\frac{1}{1-a}\epsilon '\varphi _1, \label{varphi1}\\
\delta _\xi \varphi _2 &=& \varphi _2'\epsilon -\frac{a}{1-a} \epsilon '\varphi _2 +\frac{1}{1-a}\eta \varphi _1, \label{varphi2}
\end{eqnarray}
where it can be seen that $T$ behaves like a Sugawara stress tensor and the field $S$ has the exact same behaviour as the field $\Lambda $ in (\ref{lambda_trafo}). Note that the combinations $S\varphi _1$, $S\varphi _2$ and $\varphi _1^{1-a}$ have $a$-independent conformal weights and transform like
\begin{align}
    \delta _\xi (S\varphi _1) &= \epsilon (S\varphi _1)'+\epsilon '(S\varphi _1) ,\label{hphi1}\\
    \delta _\xi (S\varphi _2) &= \epsilon (S\varphi _2)'+2\epsilon '(S\varphi _2)+\frac{1}{1-a}\eta (S\varphi _1), \label{hphi2}\\
    \delta _\xi \varphi _1^{1-a}&:=\delta _\xi \phi=\phi '\epsilon -\epsilon '\phi , \label{y}
\end{align}
where we recognize a boundary current $S\varphi _1$, a weight-2 quantity $S\varphi _2$ and a boundary vector $\phi $. These properties under diffeomorphisms make it reasonable to take these composite fields instead of $\varphi _1 ,S $ and $\varphi _2$ as part of the boundary field content. 

From observing the behaviour of $P$ and $T$ in (\ref{ads01})-(\ref{ads02}) as well as (\ref{ads_defs}) one concludes that these fields belong to the coadjoint representation of the warped Virasoro group provided we set $\eta =\sigma '$. This can also be seen by switching to a dual expression by using a Fourier mode decomposition. Everything goes exactly as in section \ref{sec:asa} though the symmetry generator  has to be taken $ Q \sim \epsilon T +\sigma P$ instead of (\ref{charge}). Note, that this choice is consistent with the relation $[\sigma ]=[\epsilon ]+1$ between the weights of transformation parameters and also correctly reproduces the transformations of the other fields $S\varphi _1$, $S\varphi _2$ and $\phi $. At the end of the day, we have
\begin{eqnarray}
i[T_n, T_m]&=&(n-m)T_{m+n},\label{alg1.1}\\
i[T_n, P_m]&=&-m\,P_{m+n}+ik(n^2-n)\delta _{m+n,0}, \label{alg1.2}\\
i[P_n, P_m]&=&\kappa n\delta _{m+n,0},\label{alg1.3}
\end{eqnarray}
where we shifted $P\to iP$, $\eta \to i\eta $ and the zero-mode $P_n\to P_n+ik\delta _{n,0}$. This last zero mode shift was performed out of convenience so that the bracket relations contain the maximal subalgebra of (\ref{akv_mode1})-(\ref{akv_mode3}). The non-zero constants $k$ and $\kappa $ describe the central extensions appearing, we however have to leave them undetermined because of the same ambiguity as in section \ref{sec:asa} where an arbitrary prefactor remained in the construction of the mode decomposition. Furthermore it has to be emphasized that it is unclear at this point whether these fields are actually related to charges or not. As an additional remark we want to point out that the change of basis $T_n\to T_n+i\frac{kn}{\kappa }P_n$ eliminates the twist term in (\ref{alg1.2}) and yields a warped Virasoro algebra 
\begin{eqnarray}
i[T_n, T_m]&=&(n-m)T_{m+n}+\frac{c}{12}(n^3-n)\delta _{n+m,0}\\
i[T_n, P_m]&=&-m\,P_{m+n}\\
i[P_n, P_m]&=&\kappa n\delta _{m+n,0}
\end{eqnarray}
where we redefined $c=-12k^2/\kappa $. It is clear from the transformation that this only works for a non-vanishing current level $\kappa $, and therefore does not apply to the algebras in the next two sections. 

Recently, a general expression of the canonical boundary charges in the second-order formulation was used to show their finiteness and integrability for a certain class of gauges \cite{Ruzziconi:2020wrb}. This however required some amount of machinery already for simple cases like the JT-model. Also, a crucial ingredient there was the use of Sachs--Bondi gauge with the additional requirement of a linear dilaton. As in our case the dilaton is not a linear function (cf.\ (\ref{dil_untransformed})) the computations got much more involved, especially because the equations of motion do not simplify a lot. It therefore still remains an open problem to find a finite and integrable charge expression for general models of the a-family.
It is however interesting to see that the above result is an extension of \cite{Godet:2020xpk} where for the case of the JT-model an explicit mapping between the phase spaces in Sachs--Bondi and Fefferman--Graham gauge is provided showing that the asymptotic symmetries are reduced in the latter case. Similar things happen here if one takes the above charge expressions seriously: Comparison with the symmetries in section \ref{sec:asa} indeed leads back to a single Virasoro algebra.



\subsection{Minkowski ground state models}
Line elements that asymptote to Minkowski space are reached by the condition $a=b+1$. An interesting class among these are spherically reduced models from $D>3$ for which the parameter $a$ takes the values
\begin{align}
    a=\frac{D-3}{D-2} \qquad \rightarrow \qquad a\in [1/2,1 ).
\end{align}
The radial coordinate $r$ is again defined by (\ref{phi_r}) and under an appropriate choice of $B=2(1+b)=2a$ the line element takes the form 
\begin{eqnarray}
\dd s^2=2i\dd u \dd r + \Big  [1+2\mathcal{C}[(1-a)r]^{-\frac{a}{1-a}} \Big ]\dd u^2,
\end{eqnarray}
where the asymptotic region $R\to 0$ is reached by $r\to \infty $. Letting the allowed diffeomorphisms act on the line element and the dilaton yields
\begin{eqnarray}
\dd s^2&=&2i\dd u \dd r + \Big  [P(u)r+T(u)+S(u)r^{-\frac{a}{1-a}}+\mathcal{O}(r^{-\frac{1}{1-a}}) \Big ]\dd u^2,\\
\Phi (u,r) &=& \varphi _1(u)r^{\frac{1}{1-a}}+\varphi _2(u)r^{\frac{a}{1-a}} + \mathcal{O}(r^{\frac{2a-1}{1-a}}),
\end{eqnarray}
where expressions in front of the different orders in $r$ have been given new names
\begin{subequations}
\begin{align}
P(u) &:= -2i\frac{f''}{f'}, & \varphi _1(u) &:= \Big (\frac{1-a}{f'}\Big )^{\frac{1}{1-a}},\\
T(u) &:= 2i\Big (g''-g'\frac{f''}{f'}\Big ), & \varphi _2(u) &:= \Big (\frac{1-a}{f'}\Big )^{\frac{1}{1-a}} \frac{g'}{1-a} = \varphi _1\frac{g'}{1-a},\\
S(u) &:= 2\mathcal{C}(1-a)^{\frac{a}{a-1}}f'^{\frac{2-a}{1-a}} .
\end{align}
\end{subequations}
Again, it can be seen that only two of these fields are independent from each other which is a consequence of the radial equations of motion. To not further restrict the AKV's the last part of the boundary conditions has to be chosen as
\begin{align}
   \delta g_{uu} &= \delta P(u)r+\delta T(u)+\delta S(u)r^{-\frac{a}{1-a}} + \mathcal{O} (r^{-\frac{1}{1-a}}), \\
    \delta \Phi &= \delta \varphi _1(u)r^{\frac{1}{1-a}}+\delta \varphi _2(u)r^{\frac{a}{1-a}} + \mathcal{O} (r^{\frac{2a-1}{1-a}}) ,
\end{align}
which includes $\mathcal{O}(r)$ terms in the metric. This does not spoil our asymptotic behaviour as the Ricci-scalar is proportional to the second radial derivative of $g_{uu}$ and thus still vanishes asymptotically. The asymptotic Killing vectors (\ref{AKVs}) induce the transformations
\begin{eqnarray}
\delta _\xi P&=&\epsilon P'+\epsilon 'P-2\epsilon '' ,\\
\delta _\xi T&=& \epsilon T'+2\epsilon 'T-2\eta '-\eta P,\\
\delta _\xi (S\varphi _1) &=& \epsilon (S\varphi _1)'+\epsilon '(S\varphi _1) ,\\
\delta _\xi (S\varphi _2) &=& \epsilon (S\varphi _2)'+2\epsilon '(S\varphi _2)+\frac{1}{1-a}\eta (S\varphi _1), \\
\delta _\xi \varphi _1^{1-a}&:=&\delta _\xi \phi=\phi '\epsilon -\epsilon '\phi ,
\end{eqnarray}
where again the redefinitions $\eta \rightarrow i\eta $ and $P\rightarrow iP$ were made. Also, this time we directly listed the composite fields $S\varphi _1$, $S\varphi _2$ and $\phi $. Their constituents transform just in the same way as in (\ref{ads03})-(\ref{varphi2}).

Switching to Fourier modes for the fields $P$ and $T$ with an again shifted zero mode $P_n \to P_n+ik\delta _{n,0}$ we arrive at the to some readers more familiar expression
\begin{eqnarray}
i[T_n,T_m]&=&(n-m)T_{m+n},\label{alg_mink.1}\\
i[T_n,P_m]&=&-mP_{m+n}+ik(n^2-n)\delta _{m+n,0},\label{alg_mink.2}\\
i[P_n,P_m]&=&0,\label{alg_mink.3}
\end{eqnarray}
where the scaling of the generating charges like in the previous subsection again remains undetermined. This is a twisted warped conformal algebra regardless of the value of $D$, the twist term cannot be brought to vanish because of the vanishing current level in (\ref{alg_mink.3}). Although these charges consistently reproduce the transformation behaviour of our boundary fields it is again not clear whether $P$ and $T$ are really related to charges. We have to leave this point open for now. 

\subsection{Rindler ground state models}
Rindler ground state models are characterized by $b=0$. For the Ricci scalar (\ref{Rab}) to vanish at $\Phi \to \infty $ the restriction $a<1$ is made. In principle one could also allow for values $1<a<2$ which however shall not be done here. Transforming to the new radial coordinate (\ref{phi_r}) leads to $r\to \infty $ as the asymptotic region. The line element and dilaton for this model then take the form
\begin{subequations}
\label{rindler_metric_dil}
\begin{eqnarray}
    \dd s^2&=&2i\dd u \dd r+\Big [(1-a)r+2\mathcal{C} [(1-a)r]^{-\frac{a}{1-a}}\Big ]\dd u^2,\\
    \Phi &=&[(1-a)r]^{\frac{1}{1-a}}.
\end{eqnarray}
\end{subequations}
Like before, we want to take(\ref{AKVs} ) as allowed transformations and therefore need to define an appropriate boundary condition for $g_{uu}$. Acting with (\ref{finite_trafo} ) on the fields leads then to
\begin{align}
    ds^2&=2i\dd u\dd r+\Big [P(u)r+T(u)+S(u)r^{-\frac{a}{1-a}}+\mathcal{O}(r^{-\frac{1}{1-a}})\Big ]\dd u^2,\\
    \Phi (u,r)&=\varphi _1(u)r^{\frac{1}{1-a}}+\varphi _2r^{\frac{a}{1-a}}+\mathcal{O}(r^{\frac{2a-1}{1-a}})
\end{align}
with the definitions
\begin{subequations}
\begin{align}
P(u) &:= (1-a)f'-2i\frac{f''}{f'}, & \varphi _1(u) &:= \Big (\frac{1-a}{f'}\Big )^{\frac{1}{1-a}},\\
T(u) &:= (1-a)f' g'+2i\Big (g''-g'\frac{f''}{f'}\Big ), & \varphi _2(u) &:= \Big (\frac{1-a}{f'}\Big )^{\frac{1}{1-a}} \frac{g'}{1-a} = \varphi _1\frac{g'}{1-a},\\
S(u) &:= 2\mathcal{C}(1-a)^{\frac{a}{a-1}}f'^{\frac{2-a}{1-a}}
\end{align}
\end{subequations}
of the new fields in terms of the generating functions $f(u),g(u)$. We therefore set the last part of the boundary conditions to
\begin{align}
    \delta g_{uu}=\delta P(u)r+\delta T(u)+\delta S(u)r^{-\frac{a}{1-a}}+\mathcal{O}(r^{-\frac{1}{1-a}}).
\end{align}
The induced transformations of the boundary fields are
\begin{eqnarray}
\delta _{\xi }P&=&\epsilon P'+\epsilon 'P-2\epsilon '' ,\label{rind.1}\\
\delta _{\xi }T&=&\epsilon T'+2\epsilon 'T-2\eta '-\eta P,\label{rind.2}\\
\delta _\xi (S\varphi _1) &=& \epsilon (S\varphi _1)'+\epsilon '(S\varphi _1), \\
\delta _\xi (S\varphi _2) &=& \epsilon (S\varphi _2)'+2\epsilon '(S\varphi _2)+\frac{1}{1-a}\eta (S\varphi _1), \\
\delta _\xi \varphi _1^{1-a}&:=&\delta _\xi \phi=\phi '\epsilon -\epsilon '\phi 
\end{eqnarray}
with $\eta \rightarrow i\eta $ and $P \rightarrow iP$. Switching to a Fourier representation the bracket relations between $P$ and $T$ read
\begin{eqnarray}
i[T_n, T_m]&=&(n-m)T_{m+n},\label{alg_rindler.1}\\
i[T_n, P_m]&=&-m\,P_{m+n}+ik(n^2-n)\delta _{m+n,0},\label{alg_rindler.2}\\
i[P_n, P_m]&=&0,\label{alg_rindler.3}
\end{eqnarray}
which is again a twisted warped conformal algebra. The same considerations as in the previous subsection apply. 

\section{Conclusions}\label{sec:concl}
Let us briefly summarize our results and discuss their implications. For a one-parameter family of asymptotically AdS$_2$ dilaton gravities we found a set of asymptotic conditions and a boundary action leading to a consistent variational problem. Besides, the action was finite on-shell. We found an asymptotic Virasoro symmetry in these models. Even without explicit boundary charges (we leave this cumbersome construction for a future work), we see compelling evidence that these gravity models have some CFT$_1$ theories as holographic duals. By a conformal redefinition of the metric we extended these results to a huge set of other dilaton gravities (all having rather unusual asymptotic geometries). We also checked the consistency of entropy formulae on the gravity side. 

Regarding possible CFT$_1$ duals, our results are less definite. We found a Schwarzian action though it describes (a part of) the equations of motion rather than the boundary action of the gravity model with our asymptotic variables. We checked that the gravity side entropy is not consistent with the usual Cardy formula though it is consistent with a modified Cardy-like formula. In other words, at the CFT side we see similarities to the JT case, but there are also notable differences. This may indicate, that the holographic duals to the theories considered here are some SYK-like models rather than the SYK model itself.

Finally, we considered an even wider class of models and more general asymptotic conditions. In these cases, we only found possible asymptotic symmetry algebras without going into analyzing boundary actions and the consistency of the variational problem. They appeared to be various twisted warped conformal algebras.  
\acknowledgments
We thank Daniel Grumiller for collaboration in the initial phase
and for remarks on earlier versions of this paper. D.V. was supported in parts by the
S\~ao Paulo Research Foundation FAPESP, project 2016/03319-6, by the grant 305594/2019-2 of
CNPq, and by the Tomsk State University Competitiveness
Improvement Program. FE was supported by project P 33789 of the
Austrian Science Fund (FWF). C.V. thanks PPGFIS and UFBA for their Visiting Professor program. 

\appendix
\section{Exact solutions of the equations of motion}\label{sec:ex}
In the Euclidean signature, the equations of motion of dilaton gravity are most easily solved \cite{Bergamin:2004pn} in complex variables
\begin{equation}
\label{D04}
Y\equiv\frac{1}{\sqrt{2}}\left(Y^{1}+iY^{2}\right),\;\;\;\bar Y\equiv\frac{1}{\sqrt{2}}\left(Y^{1}-iY^{2}\right),\;\;\;
e\equiv\frac{1}{\sqrt{2}}\left(e_{1}+ie_{2}\right),\;\;\;\bar e\equiv\frac{1}{\sqrt{2}}\left(e_{1}-ie_{2}\right).
\end{equation}
In these variables, the first-order action \eqref{D02} becomes
\begin{equation}\label{D05}
I_{\mathrm{1st}}=\int_{\mathcal M} \left[\bar{Y}\mathrm{d}e+Y\mathrm{d}\bar{e}+\Phi\mathrm{d}\omega-i\bar{Y}\omega\wedge e+iY\omega\wedge\bar{e}+i\mathcal{V}\bar{e}\wedge e\right].
\end{equation}
In terms of the complex variables
\begin{equation}
\mathcal{V}=U(\Phi)\, \bar Y Y + V(\Phi).
\end{equation}
The equations of motion for this action read
\begin{eqnarray}
\mathrm{d}\Phi+iY\bar{e}-i\bar{Y}e	&=&	0,\label{S01a}\\
\mathrm{d}Y+i\mathcal{V}e-iY\omega	&=&	0,\\
\mathrm{d}e-i\omega\wedge e+i\left(\partial_{\bar{Y}}\mathcal{V}\right)\bar{e}\wedge e	&=&	0,\\
\mathrm{d}\omega+i\partial_{\Phi}\mathcal{V}\bar{e}\wedge e	&=&	0.\label{S01d}
\end{eqnarray}
Classical solutions with a non-constant dilaton\footnote{In this paper we do not consider the constant dilaton solutions since they lead to trivial holographic duals \cite{Grumiller:2015vaa}.} are expressed through two functions of the dilaton,
\begin{equation}\label{S03}
Q\left(\Phi\right)=\int^\Phi \mathrm{d}y\; U\left(y\right),\;\;\;\;\;\;w\left(\Phi\right)=\int^\Phi \mathrm{d}y\; V\left(y\right)e^{Q\left(y\right)}.
\end{equation}
One can show that
\begin{equation}\label{S02}
\mathcal C \equiv w\left(\Phi\right) + Y\bar{Y} e^{Q\left(\Phi\right)}
\end{equation}
is a constant of motion, meaning that $\mathrm{d}\mathcal{C}=0$ on shell. The solutions of equations of motion can be expressed as follows \cite{Bergamin:2004pn}
\begin{equation}\label{S04}
e = Y e^{Q}\mathrm{d}f,\;\;\;\;\;\;\;\bar e = i\frac{\mathrm{d}\Phi}{Y}+\bar Y e^{Q}\mathrm{d}f,\;\;\;\;\;\;\;\omega=\mathcal{V}e^Q\mathrm{d}f-i\frac{\mathrm{dY}}{Y}.
\end{equation}
Here $f$ is a complex function. By requesting that $e$ equals to the complex conjugate of $\bar{e}$ one fixes 
\begin{equation*}
\Im \mathrm{d} f= - \frac{e^{-Q(\Phi)}}{2|Y|^2} \mathrm{d}\Phi
\end{equation*}
while $\theta\equiv \Re f$ remains arbitrary. Let us write $Y=|Y|e^{i\alpha}$. The absolute value $|Y|$ can be expressed through $\Phi$ and $\mathcal{C}$ by means of Eq.\ (\ref{S02}),
\begin{equation}
|Y|^2=[\mathcal{C}-w(\Phi)]e^{-Q(\Phi)}. \label{absY}
\end{equation} 
Hence,
\begin{equation}
\omega=\mathcal{V}e^Q\, \dd\theta +\dd\alpha,\qquad
e=|Y|e^{i\alpha}e^Q\, \dd\theta - \frac{ie^{i\alpha} \, \dd \Phi}{2|Y|}
\end{equation}
while the line element becomes
\begin{equation}\label{S05}
\mathrm{d}s^2 = e^Q \left[\xi\mathrm{d}\theta^2+\xi^{-1}\mathrm{d}\Phi^2\right],\;\;\;\;\;\xi \left(\Phi\right) \equiv 2\left[\mathcal{C} - w(\Phi)\right].
\end{equation}
The solution depends on one constant $\mathcal{C}$ (which roughly corresponds to the black hole mass) and 3 arbitrary functions $\Phi$, $\xi$, and $\theta$. The latter correspond to 3 continuous gauge symmetries of dilaton gravity in the first-order formulation. We stress, that no gauge fixing conditions were imposed so far.

One can derive the following on-shell identities
\begin{eqnarray}
&& \Phi \dd\omega = \Phi \frac {\partial}{\partial\Phi}
\left( \mathcal{V}(\Phi) e^Q \right) \dd \Phi\wedge \dd \theta \,,
\label{id2}\\
&&\bar Y (\dd e-i\omega\wedge e)= \bar YYU(\Phi) e^Q\, \dd \Phi\wedge \dd \theta \,,\label{id3}\\
&& i \bar e \wedge e = - e^Q\, \dd \Phi\wedge \dd \theta\,.\label{id4}
\end{eqnarray}
Before computing the $\Phi$-derivative in (\ref{id2}) one has to express $\bar YY$ through $\Phi$ with the help of (\ref{absY}).
These identities allow to compute the on-shell action. It vanishes for for all dilaton gravity models with a linear $V(\Phi)$ and an inverse linear $U(\Phi)$. In particular, for the $a$-family,
\begin{equation}
    \left. I_{\mathrm{1st}} \right|_{\mathrm{on-shell}}=0 .\label{I0}
\end{equation}

For the potentials (\ref{ab01}), we have
\begin{equation}
e^Q=\Phi^{-a},\qquad w=-\frac B{2(2-a)} \Phi^{b+1}.\label{eQw}
\end{equation}
The curvature scalar is given by the formula \cite{Bergamin:2004pn}
\begin{equation}
    R=-2a\mathcal{C}\Phi^{a-2} -\frac{Bb(b-a+1)}{b+1}\Phi^{b+a-1}.\label{Rab}
\end{equation}

Let us consider the case $b=1-a$.
If $\mathcal{C}<0$ the metric (\ref{S05}) has a Euclidean horizon (a tip) for $\Phi=\Phi_0$ with
\begin{equation}
\Phi_0^{2-a}=-2\mathcal{C}. \label{Phi0}
\end{equation}
(If $\mathcal{C}>0$ the metric has a naked singularity and corresponds to a negative mass black hole.) To avoid a conical singularity, $\theta$ should have the period
\begin{equation}
\beta_\theta=\frac{4\pi}{(2-a)\Phi_0^{1-a}}.\label{betat}
\end{equation}

\bibliographystyle{fullsort.bst}
 
\bibliography{review,additions} 

\end{document}